\documentclass[sigconf]{acmart}
\AtBeginDocument{%
  }

\copyrightyear{2026}
\acmYear{2026}
\setcopyright{cc}
\setcctype{by}
\acmConference[ASSETS '26]{The 28th International ACM SIGACCESS Conference on Computers and Accessibility}{October 25--28, 2026}{Vila Nova de Gaia, Portugal}
\acmBooktitle{The 28th International ACM SIGACCESS Conference on Computers and Accessibility (ASSETS '26), October 25--28, 2026, Vila Nova de Gaia, Portugal}
\acmDOI{10.1145/3797867.3828978}
\acmISBN{979-8-4007-2521-0/2026/10}

\begin{document}

\title[Bigger than the EAR BOX]{Bigger than the EAR BOX: A Theory-Grounded Review of XR Accessibility Research for Deaf and Hard of Hearing Communities}

\author{Shuxu Huffman}
\affiliation{%
  \institution{Gallaudet University}
  \city{Washington}
  \state{District of Columbia}
  \country{USA}
}
\email{shuxu.huffman@gallaudet.edu}

\author{Michaela Okosi}
\affiliation{%
  \institution{Gallaudet University}
  \city{Washington}
  \state{District of Columbia}
  \country{USA}
}
\email{Michaela.okosi@gallaudet.edu}

\author{Jose Merino}
\affiliation{%
  \institution{Accessible Human-Centered Computing}
  \institution{Gallaudet University}
  \city{Washington}
  \state{District of Columbia}
  \country{USA}
}
\email{joe.merino@gallaudet.edu}

\author{Tifanie Bouchara}
\affiliation{%
  \institution{LISN, VENISE Team}
  \institution{Université Paris-Saclay, CNRS}
  \city{Orsay}
  \country{France}
}
\email{tifanie.bouchara@universite-paris-saclay.fr}

\author{Abraham Glasser}
\affiliation{%
  \institution{Science, Technology, Accessibility, Mathematics, and Public Health (STAMP)}
  \institution{Gallaudet University}
  \city{Washington}
  \state{District of Columbia}
  \country{USA}
}
\email{abraham.glasser@gallaudet.edu}

\author{Christian Vogler}
\affiliation{%
  \institution{Technology Access Program}
  \institution{Gallaudet University}
  \city{Washington}
  \state{District of Columbia}
  \country{USA}
}
\email{christian.vogler@gallaudet.edu}

\author{Raja Kushalnagar}
\affiliation{%
  \institution{Gallaudet University}
  \city{Washington}
  \state{District of Columbia}
  \country{USA}
}
\email{raja.kushalnagar@gallaudet.edu}


\renewcommand{\shortauthors}{Huffman et al.}

\begin{abstract}
    Extended Reality (XR) technologies have received growing attention in accessibility research involving Deaf and Hard of Hearing (DHH) communities. Yet less attention has been given to the assumptions shaping this work. We present a theory-grounded review of an ACM-published corpus of XR research involving DHH users. Drawing on Disability Studies, Deaf Studies, and DeafSpace, we develop a theoretical framework with four analytical dimensions: orientation toward access, distribution of responsibility, conceptualization of DHH communities, and spatial and perceptual assumptions. We apply this framework to 53 XR studies involving DHH users, identified through searching and screening of ACM publications from 2015 to 2025. Our analysis shows that XR accessibility is frequently framed as supporting communication within hearing-default environments, while overlooking the diversity of DHH communities. We identify directions for redistributing accessibility labor and reconfiguring space, and offer our framework as a tool for future XR research involving DHH communities.
\end{abstract}

\ccsdesc[500]{Human-centered computing~Accessibility theory, concepts and paradigms}
\ccsdesc[300]{Social and professional topics~People with disabilities}

\keywords{Accessibility, Deaf and hard of hearing, literature review, Extended Reality}

\received{22 April 2026}

\maketitle

\section{Introduction}

Extended Reality (XR) technologies, including virtual reality, augmented reality, and mixed reality, have increasingly been explored in accessibility research within Human-Computer Interaction (HCI) over the past decade~\cite{mott_accessible_2019, gerling_critical_2021, dudley2023inclusive}. Among these efforts, Deaf and Hard of Hearing (DHH)\footnote{We use ``DHH'' as a broad umbrella term referring to heterogeneous groups whose relationships to hearing status, language, technology use, and community membership are not interchangeable. This includes, for example, Deaf signers, hard of hearing individuals, late-deafened adults, and deaf individuals who primarily use spoken language. By contrast, when we refer to the ``Deaf community'' with a capital D, we more specifically refer to communities organized around sign language and shared cultural and linguistic practices. We note that these boundaries are neither fixed nor exhaustive, and that individuals may identify with, move across, or relate to these categories in complex ways.} communities have become a recurring focus. 

Prior work has shown that DHH-related accessibility research is often concentrated around communication technologies, such as real-time captioning or signing avatar systems~\cite{mack2021we}, suggesting that such work tends to follow particular ways of conceptualizing access. This raises the question of whether these patterns are reproduced or reconfigured in XR research, where interaction unfolds in immersive environments and spatial organization plays a central role, motivating an examination of how XR research understands access, users, and space.

Accessibility research in HCI has been shaped by interdisciplinary perspectives, drawing on multiple fields to understand how technologies define users and forms of access~\cite{mankoff_disability_2010, mack2021we}. Within this broader conversation, \textit{Disability Studies} provides critical tools for understanding how disability is conceived, produced, and designed for~\cite{bennett2018interdependence, kafer2013feminist, mills2010deaf}. In 2010, Mankoff et al. first argue that it should serve as a central theoretical foundation for accessibility research~\cite{mankoff_disability_2010}. Because our focus is on XR research involving DHH communities, it is also essential to engage with \textit{Deaf Studies}, which examines language, culture, and technology in relation to Deaf communities, including how technologies may reinforce or challenge existing power relations~\cite{forshay2016signaloud, kusters2017innovations}. Finally, XR is a spatial medium, requiring attention to how interaction unfolds across immersive space. We therefore draw on \textit{DeafSpace}, a Deaf-centered architectural framework that foregrounds visual attention, spatial orientation, and shared fields of interaction~\cite{bauman2024deafspace}. In this paper, we bring these three bodies of scholarship together to review $53$ ACM-published XR research papers involving DHH populations from 2015 to 2025. We treat this bounded corpus as an analytically meaningful site for studying how HCI, accessibility, and interactive computing communities have conceptualized DHH access in XR.

This paper makes three contributions. First, we introduce a theory-grounded analytical framework that draws together \textit{Disability Studies}, \textit{Deaf Studies}, and \textit{DeafSpace}, and derive a set of analytical dimensions for examining XR research involving DHH communities. Second, we apply this framework to a decade of XR research, providing a structured analysis of $53$ papers and identifying recurring patterns in how access, DHH communities, and space are conceptualized. Third, we reflect on these patterns to surface key tensions in current approaches and outline directions for reimagining XR accessibility, including moving beyond deficit-oriented framings, redistributing accessibility labor across users, technologies, and ecosystems, and leveraging XR's spatial affordances to support Deaf-centered interaction.

\section{Related Work}
\label{rw:litReview}

\subsection{Descriptive Reviews of Accessibility Research}

Literature reviews play an important role in accessibility research by summarizing field-level trends, identifying common questions and methods, and showing how research areas have developed over time~\cite{mack2021we, mcdonnell2024envisioning}. In 2021, Mack et al. conducted the first comprehensive review of accessibility papers published at CHI and ASSETS over a 26-year period~\cite{mack2021we}. Their analysis showed that only 11.3\% of papers focused on DHH users, compared to 43.5\% addressing blind and low vision (BLV) users. They further reported that nearly two thirds of DHH papers (64.9\% of 57 total) focused on communication related technologies. 

Beyond such field-level overviews, researchers have conducted literature reviews centered on specific populations or domains. These include reviews focused on BLV users~\cite{bhowmick2017insight, brule2020review, grussenmeyer2017accessible}, autism~\cite{pennisi2016autism, spiel2019agency, rizvi2024robots}, ADHD~\cite{spiel2022adhd}, children with special needs~\cite{baykal2020collaborative}, and DHH users~\cite{mcdonnell2024envisioning}. Other reviews have focused on methodologies~\cite{brule2020review} or specific technologies~\cite{grussenmeyer2017accessible, lorah2015systematic, alonzo2025review}, including XR applications in education for BLV users~\cite{hamash2024breaking} as well as XR accessibility more broadly~\cite{dudley2023inclusive}. 
 
Collectively, these works document dominant areas of focus and methodological patterns, highlighting which topics and populations receive attention and which remain under-examined. At the same time, they leave room for other kinds of questions, including how access, users, and interaction are being understood in XR research, and how those conceptualizations shape the field's directions.

\subsection{Critical and Theory-Grounded Reviews}

In addition to descriptive synthesis, prior work has applied structured and critical lenses to interrogate the assumptions underlying accessibility research. Vines et al. critically analyzed 30 years of SIGCHI aging research and demonstrated that the literature largely frames aging as a problem to be managed by technology, conflates aging with accessibility, and reproduces deficit-oriented stereotype~\cite{vines2015age}. Rizvi et al. critically examined 142 human robot interaction studies on autism, showing that many reproduce ableist assumptions and systematically exclude autistic perspectives and offering recommendations for more inclusive and equitable future research~\cite{rizvi2024robots}. Spiel et al. reviewed accessibility technologies for autistic children~\cite{spiel2019agency} and for people with ADHD~\cite{spiel2022adhd}, finding that this work largely reflects neurotypical, deficit-based assumptions and marginalizes neurodivergent agency, and they proposed directions for more inclusive research and design practices. 

Related work has also used prior reviews to inform design guidance, as in Bauer et al.'s XR guidelines for autism interventions~\cite{bauer2023extended}. McDonnell and Findlater introduced a theory-grounded analytical lens combining \textit{Disability Studies}, \textit{Deaf Studies}, \textit{Disability Justice}, and \textit{Communication Studies} to examine captioning literature~\cite{mcdonnell2024envisioning}. Gerling et al. critically examined how accessibility is conceptualized in virtual reality research and identified broader patterns in the field~\cite{gerling2025equitable}. Together, these theory-grounded and critically oriented literature reviews demonstrate how analytical frameworks can be used to surface implicit assumptions, power relations, and exclusions within accessibility research.

However, such approaches have not been systematically applied to XR research involving DHH communities. Our work fills this gap by developing a theoretical framework grounded in \textit{Disability Studies}, \textit{Deaf Studies}, and \textit{DeafSpace}, and applying it to a decade of XR accessibility research involving DHH users. In doing so, we offer a lens for critically interpreting prior work and for envisioning alternative, Deaf-centered futures in immersive technologies.

\section{Theoretical Framework}

In this section, we introduce our theoretical framework and the analytical dimensions derived from it. We draw on three complementary bodies of scholarship---\textit{Disability Studies}, \textit{Deaf Studies}, and \textit{DeafSpace}---to develop a lens for examining how XR research conceptualizes access, users, and interaction. From these foundations, we derive four analytical dimensions that guide our interpretive review of prior XR research involving DHH communities.

\subsection{Theoretical Foundations}
\label{rw:theoretical_framework}

\begin{figure}[h]
  \centering
  \includegraphics[width=0.9\linewidth]{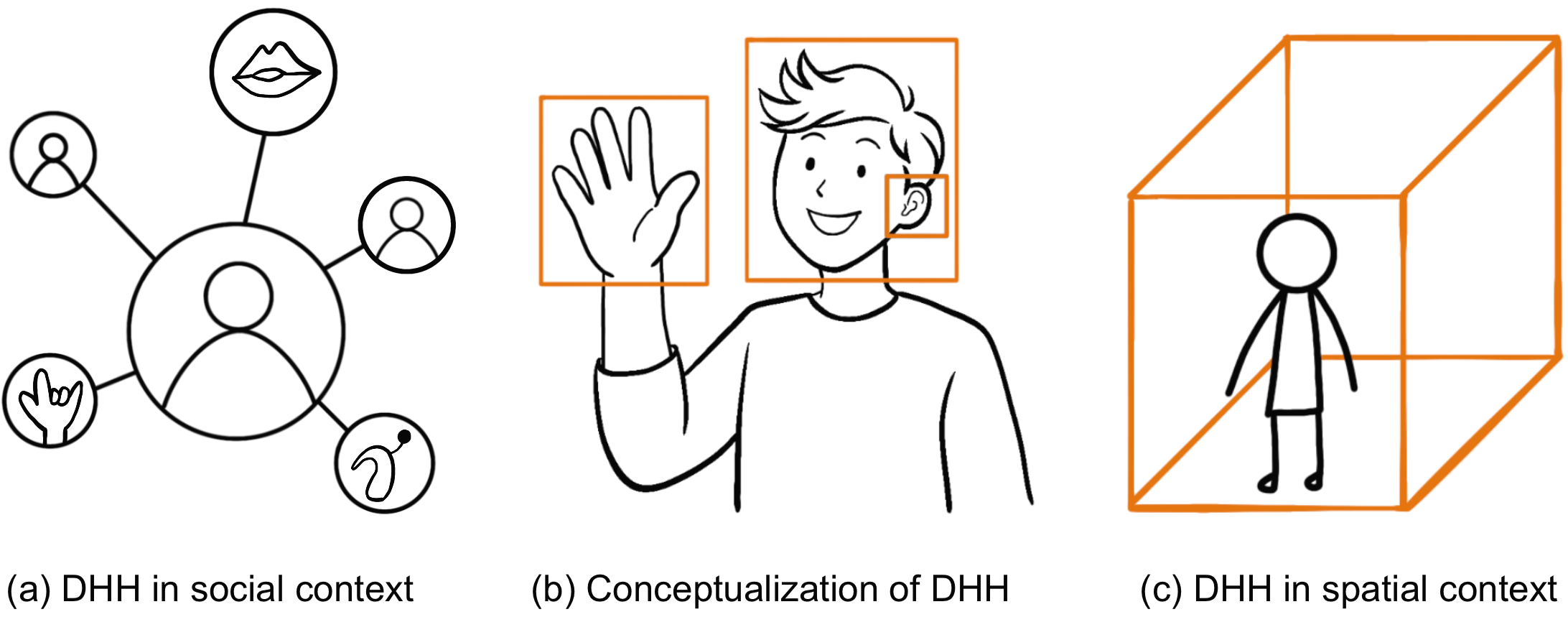}
  \caption{Illustration of how our theoretical framework brings together Disability Studies, Deaf Studies, and DeafSpace to conceptualize XR research involving DHH users and examine access. (a) DHH users are situated within broader social structures and relationships. (b) DHH users are conceptualized as composed of multiple, overlapping dimensions, including communication practices and embodied experience. (c) XR systems organize access through spatial configurations of interaction, visibility, and information.}
  \Description{Three side-by-side diagrams. The left diagram shows a central figure surrounded by connected nodes representing other people, lips, a cochlear implant, and the ILY sign, illustrating how DHH users are situated within broader social structures and relationships. The middle diagram shows a human figure divided into sections, illustrating that DHH users are composed of multiple overlapping dimensions. The right diagram shows a person inside a three-dimensional box.}
  \label{fig:framework}
\end{figure}

\subsubsection{\textbf{Disability Studies} and Access in HCI}

\textit{Disability Studies} has long challenged traditional medical models, which locate disability within impaired bodies and treat it as a problem to be corrected at the individual level~\cite{union1976fundamental, berger2013introducing, shakespeare1996disability}, and emphasized social models that understand disability as produced through environments, technologies, and institutional arrangements~\cite{oliver2012social, shakespeare1996disability, hamraie2017building, michael1989disability}. From this perspective, access is not a technical add-on but a social and relational concern shaped by design decisions and power relations, situating users within broader social structures and relationships (Fig. \ref{fig:framework}(a)).

Within HCI, \textit{Disability Studies} has informed critiques of technologies that implicitly assume normative bodies and senses, positioning disabled users as edge cases who must adapt to existing systems~\cite{mankoff_disability_2010, hamraie2017designing, bennett2018interdependence, gerling_critical_2021}. Rather than asking how technologies can ``fix'' disabled users, this work reflects how technologies themselves might be reconfigured around diverse ways of being~\cite{goering2015rethinking}. \textit{Crip technoscience} extends this perspective by foregrounding disabled people not merely as recipients of accessible design, but as epistemic contributors whose lived experiences constitute a source of design knowledge~\cite{hamraie2025crip, angelini2023criptopias, williams2019prefigurative}. More recent research further argues that accessibility is not a stable property achieved solely by removing barriers, but is instead produced through relationships among users, technologies, and contexts, as well as through situated practices of access~\cite{mack2025modeling, bennett2018interdependence}.

Prior work cautions against accessibility efforts that engage disabled people only at the level of post hoc validation~\cite{hamraie2025crip}. In such cases, disability functions primarily as a framing device to motivate innovation, while disabled people remain positioned as end users rather than contributors to problem definition, design decisions, and system making~\cite{Hamraie2023cripMaking, mills2010deaf}. When participation is limited in this way, accessibility risks becoming extractive rather than transformative~\cite{hamraie2017designing}.

These concerns are particularly salient for XR systems, which often introduce new interactional demands and redistribute cognitive and coordination labor~\cite{gerling_critical_2021, mott_accessible_2019}. From a \textit{Disability Studies} perspective, XR accessibility cannot be reduced to the presence or absence of assistive features. Instead, it calls for examining how access is framed, how responsibility for maintaining access is allocated, and the extent to which disabled users are positioned as contributors to shaping technologies.

\subsubsection{\textbf{Deaf Studies}: Language, Culture, and Group Specificity}

\textit{Deaf Studies} is an interdisciplinary field that examines Deaf people, sign languages, and Deaf communities in cultural, linguistic, social, and political terms~\cite{bauman2008open, lane2002deaf}. It foregrounds Deaf people as members of linguistically and culturally distinct communities organized around sign languages~\cite{bauman2013deaf, kusters2017innovations, padden1988deaf}. In contrast to deficit-oriented framings that treat deafness primarily as sensory \underline{loss}, \textit{Deaf Studies} emphasizes \underline{Deaf gain}, highlighting that deafness brings unique cognitive, cultural, and perceptual contributions that enrich human experience~\cite{bauman2014deaf, ladd2005deafhood, bauman2009reframing}. It recognizes sign languages as full natural languages with their own grammar, discourse practices, and sociocultural norms, rather than as subordinate to spoken languages~\cite{valli2000linguistics, stokoe2005sign, baynton1996forbidden}. From this perspective, access extends beyond compensating for hearing loss to include linguistic rights and culturally grounded forms of interaction~\cite{ladd2003understanding}, conceptualizing DHH users as composed of multiple, overlapping dimensions of communication, identity, and embodied experience (Fig. \ref{fig:framework}(b)).

The category of ``Deaf and hard of hearing'' encompasses heterogeneous populations, such as Deaf signers, hard of hearing individuals, and late-deafened adults. Life course trajectories further shape these differences, including age-related and acquired hearing loss, where many individuals have been socialized primarily in spoken language contexts and may not share the same relationships to sign language or Deaf community membership as Deaf signers~\cite{bat2000diversity, de2017buttering, ladd2003understanding}. These examples reflect a broader spectrum of relationships to Deaf culture and sign language, with communicative practices and access needs that may overlap but are \textit{not interchangeable} across contexts~\cite{bat2000diversity, parasnis1998cultural}. Prior work cautions that collapsing these groups under a single accessibility label risks obscuring meaningful differences in language use, identity, and community membership~\cite{lane1992mask, ladd2003understanding}. In technology design, such overgeneralization can lead to systems that nominally support accessibility while failing to serve the communities they claim to address~\cite{de2021good, de2017buttering}.

\textit{Deaf Studies} also informs research ethics in work involving DHH communities. Singleton et al.~\cite{singleton2014toward} report that many Deaf participants distrust hearing researchers who lack sign language fluency, as they may not fully understand participants' responses; in some cases, participants had no interpreter or accommodations and had to \textit{``figure out on his own what the researcher was trying to say.''} Harris et al.~\cite{harris2009research} further argue that ethical research requires equitable Deaf-hearing collaboration, as when hearing researchers unfamiliar with Deaf culture define Deaf people's realities, problematic assumptions can arise.

Beyond research practice, these concerns extend to the design of technologies for Deaf communities~\cite{angelini2025speculating, harris2009research, kusters2017innovations}. Technologies that enter Deaf cultural and communicative contexts, whether physical or virtual, are not neutral; they may reinforce hearing-centric norms or disrupt, and in some cases harm, established cultural and linguistic practices~\cite{forshay2016signaloud}. Design decisions may inadvertently re-center spoken language as the normative baseline and constrain Deaf users' linguistic agency~\cite{angelini2023contrasting, angelini2024bridging, baynton1996forbidden}. These framings shape assumptions about \textit{who the system is for} and \textit{whose communicative practices are prioritized}. As XR systems increasingly mediate communication and presence, \textit{Deaf Studies} calls for greater clarity regarding which populations such systems target and how design choices align with, or undermine, Deaf cultural values.

\subsubsection{\textbf{DeafSpace} and Spatializing Access in XR}
\label{rw:DeafSpace}

\textit{DeafSpace} emerged from architectural and design scholarship attentive to how Deaf ways of being shape experiences of space~\cite{edwards2014deafspace, bauman2024deafspace, cloete2025deafspace}. Rather than treating space as a neutral container, \textit{DeafSpace} emphasizes how spatial arrangements encode assumptions about attention and communication. The \textit{DeafSpace} design guidelines outline five core spatial principles: space and proximity, sensory reach, mobility and proximity, light and color, and acoustics and electromagnetic interference~\cite{gallaudetDeafSpace}. Each aimed at aligning the built environment with the ways in which Deaf people perceive, navigate, and communicate within space.

At its core, \textit{DeafSpace} challenges hearing-centric spatial norms in which auditory cues dominate at the expense of visual ones~\cite{edwards2014deafspace}. In Deaf spaces, communication is distributed across bodies, gaze, movement, and spatial orientation. Attention is dynamically negotiated through shared visual awareness and peripheral monitoring~\cite{bauman2024deafspace}. These spatial practices are integral to how Deaf communities coordinate interaction and maintain conversational flow~\cite{solvang2014accessibility}.

While \textit{DeafSpace} originated in physical architecture, its insights extend to digital and immersive environments\footnote{In this paper, we focus on the spatial and perceptual principles most directly applicable to XR interaction, such as visual attention, gaze coordination, and spatial arrangement.}. XR systems function as world-building technologies that actively construct spatial relationships, define fields of view, and organize attention~\cite{jerald2015vr, slater1997framework}. Design decisions concerning camera placement, avatar positioning, visual hierarchy, and interaction flow embed assumptions about how users see and interact with one another~\cite{madary2016real}. For Deaf users, XR environments that require frequent gaze switching~\cite{luft2023online}, fragmented visual attention~\cite{mather2012issue}, or reliance on audio-centric cues introduce additional interactional burden~\cite{rodrigues2022learning, gerling_critical_2021}. Conversely, XR spaces that support stable sightlines, shared visual fields, and negotiated gaze align more closely with Deaf spatial practices~\cite{solvang2014accessibility, bauman2024deafspace, slater1997framework}. \textit{DeafSpace} thus provides a lens for examining not only where information is placed in XR environments, but also how space itself structures access and interaction, as illustrated in Fig. \ref{fig:framework}(c).

\subsection{Analytical Dimensions}
\label{method:dimensions}

\begin{figure*}[h]
  \centering
  \includegraphics[width=\linewidth]{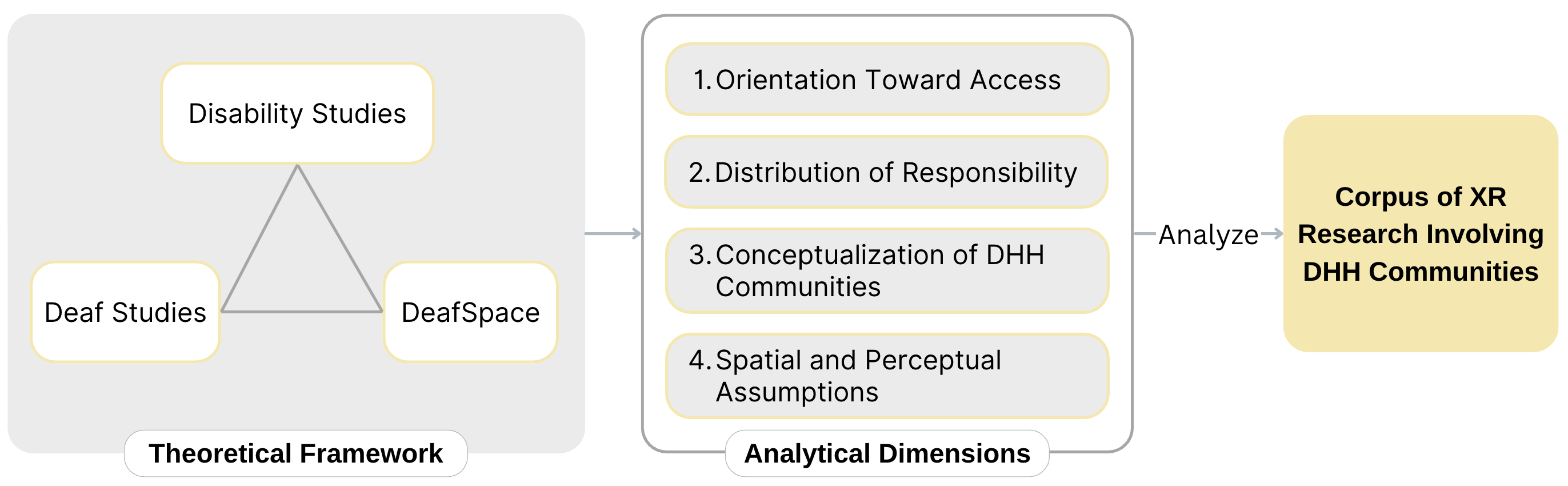}
  \caption{Overview of the analytical framework used in this study. Three theoretical foundations---Disability Studies, Deaf Studies, and DeafSpace---inform four analytical dimensions---orientation toward access, distribution of responsibility, conceptualization of DHH communities, and spatial and perceptual assumptions---that guide our interpretive analysis of a corpus of XR research involving DHH communities.}
  \Description{Diagram showing a workflow from left to right. On the left, a section labeled ``Theoretical Framework'' contains three labeled boxes---Disability Studies, Deaf Studies, and DeafSpace---connected by a triangle. An arrow points to a middle section labeled ``Analytical Dimensions,'' which lists four items: (1) Orientation Toward Access, (2) Distribution of Responsibility, (3) Conceptualization of DHH Communities, and (4) Spatial and Perceptual Assumptions. Another arrow labeled ``Analyze'' points to a box on the right that reads ``Corpus of XR Research Involving DHH Communities.''}
  \label{fig:flow}
\end{figure*}

Grounded in these theoretical foundations, we derive four analytical dimensions to guide our structured review of XR accessibility research involving DHH users (Fig. \ref{fig:flow}). Each dimension corresponds to a specific analytical question raised by our theoretical framework. \textit{Disability Studies} directs attention to how access is defined and how responsibility for access is allocated across users, technologies, and institutions. \textit{Deaf Studies} directs attention to how DHH communities are conceptualized, including whether language, culture, identity, and group specificity are made visible. \textit{DeafSpace} directs attention to how XR systems organize space, visual attention, and interaction flow. Together, these lenses allow us to examine not only what accessibility features are proposed, but also what assumptions about users, access, and space are embedded in the reviewed research. To support consistency across the research team, we operationalized these dimensions through the coding scheme shown in Table~\ref{tab:coding_scheme}.

\subsubsection{Orientation toward Access} (\textbf{\textit{Disability Studies}} and \textbf{\textit{Deaf Studies}})

This dimension examines how XR research frames access. Building on \textit{Disability Studies} and \textit{Deaf Studies}, it treats access as socially and politically constructed rather than as a neutral technical concern. Our coding scheme is informed in part by prior literature reviews of DHH accessibility research discussed in Section \ref{rw:litReview}, which identified recurring patterns such as the prominence of communication technologies.

Analytically, this dimension considers how access is conceptualized. For example, it may be understood as translation between modalities (e.g., captions or sign language interpreter overlay), sensory substitution (e.g., representing non-speech sounds through visual or haptic cues), or spatial reconfiguration (e.g., reorganizing the layout of information or interlocutors in XR space to support visual access). It also examines who is positioned as the intended beneficiary of access.

\subsubsection{Distribution of Responsibility} (\textbf{\textit{Disability Studies}})

This dimension focuses on how responsibility for establishing and maintaining access is distributed within XR systems and foregrounds forms of often invisible or normalized labor required of DHH users to make XR systems usable, whether in using XR systems or in participating in research about them.

Analytically, it examines whether access-related labor---such as setup, calibration, interpretation, or ongoing coordination---is placed primarily on DHH users or distributed across systems, other users, or institutional arrangements. It also considers whether access is supported during the research process itself, particularly in user studies involving DHH participants. In this context, appropriate accommodation refers to whether participants' communication needs were supported through measures such as direct communication in sign language, the use of interpreters, or other communication arrangements when researchers were hearing non-signers~\cite{ada_effective_communication, eud_interpreter_guidelines}.

\subsubsection{Conceptualization of DHH communities} (\textbf{\textit{Deaf Studies}})

This dimension examines how XR research defines the target population. Drawing on \textit{Deaf Studies}, it foregrounds issues of language, culture, and group specificity.

Analytically, it considers whether deafness is framed as a sensory impairment, a linguistic and cultural identity, or left unspecified, and whether diversity within DHH communities is acknowledged or treated as homogeneous. It also examines whether researchers specify the communities they design for and whether researcher positionality is reported.

\subsubsection{Spatial and Perceptual Assumptions of XR Systems} (\textbf{\textit{DeafSpace}})

This dimension focuses on the spatial and perceptual assumptions embedded in XR systems, drawing on principles from \textit{DeafSpace}. It examines how XR systems organize space, visual attention, and interaction flow.

Analytically, it considers whether these designs support shared visual access and Deaf ways of seeing and attending, or reproduce hearing-default spatial logics that prioritize audio cues or fragmented attention.

\begin{table*}[t]
\centering
\small
\caption{Analytical coding scheme used in this study.}
\begin{tabular}{p{5.3cm} p{5.2cm} p{5.5cm}}
\toprule
\textbf{Dimension and Analytical Focus} & \textbf{Coding Category} & \textbf{Possible Values} \\
\midrule

\textbf{1. Orientation toward Access} \newline
\textit{How access is framed and who it serves}
& Primary access framing
& sensory substitution; translation; spatial reconfiguration \\

&
Primary intended beneficiary
& DHH users; hearing users; mixed or ambiguous \\

&
User study involving DHH participants
& yes; no \\

\midrule

\textbf{2. Distribution of Responsibility} \newline
\textit{How access-related labor is allocated}
& Primary bearer of access labor
& user-borne; system-supported; socially or institutionally distributed; split \\

&
Appropriate accommodation
& yes; no; partially; not applicable \\

\midrule

\textbf{3. Conceptualization of DHH Users} \newline
\textit{How DHH users are defined and represented}
& Primary conceptualization
& sensory impairment; cultural or linguistic community \\

&
Target group clarity
& specified; not specified \\

&
Researcher positionality disclosure
& explicit; partial; absent \\

\midrule

\textbf{4. Spatial and Perceptual Assumptions} \newline
\textit{How space and perception are structured}
& Primary spatial framing
& hearing-centric augmentation; Deaf-centered reconfiguration \\

&
Consideration of Deaf ways of being
& explicit; implicit; absent \\

\bottomrule
\end{tabular}
\label{tab:coding_scheme}
\end{table*}

\section{Methods}

Our literature review proceeds in two stages. First, we construct a corpus of XR research involving DHH users. Second, drawing on our theoretical framework and four analytical dimensions, we conduct a structured analysis of the corpus.

\subsection{Literature Corpus Construction}

\subsubsection{Searching}


To construct the review corpus, we searched the ACM Digital Library\footnote{ACM Digital Library: \url{https://dl.acm.org/}} for peer-reviewed publications between January 2010 and January 2026 that involve XR technologies in relation to DHH users, including both full papers and shorter publication formats. Relevant XR accessibility research involving DHH communities is also published in other venues and journals, including IEEE VR, ISMAR, and other non-ACM research communities. We limited this review to the ACM Digital Library because our analysis focuses on how DHH access is conceptualized within HCI, accessibility, and interactive computing research, where ACM publications provide a bounded and high-density corpus for this topic.

We selected this search window to provide broad coverage of XR research in HCI, capturing both early exploratory work and more recent developments. Although the search window spans sixteen years, the resulting corpus effectively reflects about a decade of research. Within our corpus, eligible papers primarily appeared between 2015 and 2025\footnote{This may partly reflect the broader timeline of modern XR development, including the Oculus Rift Kickstarter in 2012 and the wider release of consumer VR systems in the mid-2010s~\cite{unity_xr_history_2021}.}, and no eligible 2026 papers were available in the ACM Digital Library at the time of the search.

We queried using combinations of terms related to Deafness and DHH populations---``deaf,'' ``hard of hearing,'' ``DHH,'' ``sign language,'' ``signer,'' and ``hearing impaired\footnote{Although the term ``hearing impaired'' is now considered outdated~\cite{NAD_ASL_FAQ}, we included it to ensure coverage of earlier work.}''---and XR technologies---``extended reality,'' ``virtual reality,'' ``augmented reality,'' ``smart glasses,'' and ``head mounted display''. We also included early-stage publications such as demos, posters, and short papers (e.g., Late Breaking Work), as these formats often introduce emerging ideas and design directions before full papers appear. This process resulted in an initial corpus of $206$ papers.


\subsubsection{Screening}

Three authors independently screened the initial corpus. Each author reviewed approximately one-third of the papers based on titles and abstracts, as well as the keywords provided in those papers. Papers were retained if they involved XR technologies and engaged DHH users or Deaf related communication practices. We then rotated assignments such that each paper was reviewed by at least two authors. Disagreements regarding inclusion were flagged and then discussed and reconciled during weekly meetings. After this initial screening stage, $74$ papers were retained. 

During the subsequent data analysis phase, we further excluded $21$ papers, resulting in a final corpus of $53$ papers\footnote{The final set of included papers is provided in the supplementary materials.}. Papers were removed for several reasons. In some cases, a full reading revealed that the work did not substantially involve XR technologies or DHH users. In other cases, multiple publications reported the same project, such as a poster or demo accompanying a later full paper (e.g.~\cite{caoSoundModVRSoundModifications2024, caoSoundModVRSoundModifications2024a, caoSupportingSoundAccessibility2024}). When this occurred, we retained the most complete publication and excluded the earlier work in progress version.

\subsection{Analytical Procedure}

Our analysis was primarily deductive and theory-informed, following prior critical and theory-grounded review approaches in HCI and accessibility research~\cite{vines2015age, mcdonnell2024envisioning}. The four analytical dimensions introduced in Section~\ref{method:dimensions} were derived from our theoretical framework before corpus analysis and used to guide a structured reading of the literature. We used these dimensions as sensitizing concepts for examining how access, responsibility, DHH communities, and space were conceptualized across the corpus. To maintain consistency across the team, we used the coding scheme shown in Table~\ref{tab:coding_scheme}, while also documenting observations that extended beyond the predefined categories. These notes informed our later synthesis of patterns across the corpus.

To align interpretations across the team, three authors independently coded an initial set of four papers using the proposed coding scheme. During this calibration phase, we discussed ambiguities in the coding categories and refined their interpretation. We then calculated Krippendorff's Alpha inter-rater reliability score~\cite{krippendorff2018content} as an indicator of agreement in applying the coding scheme, yielding a value of $0.795$, indicating substantial agreement among coders. Following this calibration, the remaining papers were divided among the three authors for initial coding. Each author subsequently reviewed a subset of 5--8 papers coded by another author to identify ambiguities, raise comments, and support alignment across the team. 

Throughout the analysis process, authors documented observations and emerging reflections. We held weekly meetings to discuss disagreements, resolve uncertainties, and align interpretations. In the later stage of analysis, we collectively reviewed our notes, identified patterns across the corpus, and developed the findings through iterative discussion and synthesis.

\subsection{Team Positionality}

Our team brings diverse lived experiences and scholarly backgrounds in Deafness, XR, and HCI. The team is primarily composed of DHH researchers, including graduate students and senior faculty with long standing experience in accessibility and XR research, with one hearing collaborator. Team members communicate primarily through ASL and written English, while also bringing diverse linguistic and cultural backgrounds, including Chinese, Spanish, French, and German. We intentionally approach this work from Deaf perspectives throughout the research process, and our positionality shapes both our interpretations and the direction of this work. Our team composition reflects a commitment to Deaf leadership in accessibility and to conducting scholarship that remains accountable to DHH communities.

\section{Results}

\begin{figure*}[h]
  \centering
  \includegraphics[width=\linewidth]{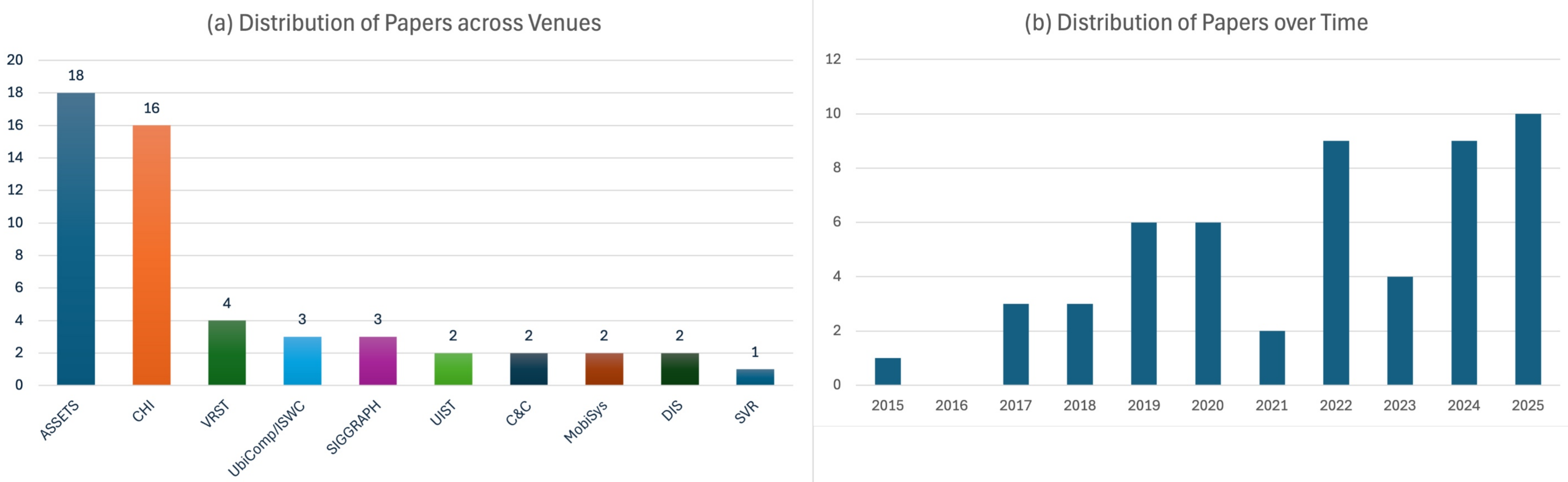}
  \caption{Distribution of papers across venues (left) and number of papers per year (right). ASSETS and CHI account for the majority of publications in the corpus, while other venues contribute smaller numbers of papers. The temporal distribution shows an increase in research activity over time, with publications growing from a single paper in 2015 to a peak in recent years.}
  \Description{The figure contains two bar charts. The left chart shows the distribution of 53 papers across venues, with ASSETS and CHI having the highest counts, followed by smaller contributions from venues such as VRST, UbiComp/ISWC, SIGGRAPH, UIST, CCC, MobiSys, DIS, and SVR. The right chart shows the number of papers published per year, increasing over time from one paper in 2015 to ten papers in 2025, with a generally upward trend indicated by a dashed line.}
  \label{fig:stat}
\end{figure*}

In our final corpus, the largest venues are ASSETS ($34.0\%$, $18/53$) and CHI ($30.2\%$, $16/53$), together accounting for nearly two thirds of the dataset (Fig. \ref{fig:stat}(a)). Examining publication trends over time, we observe an increase in research activity over the past decade, with no papers prior to 2015, followed by a steady rise from a single paper in 2015 to ten papers in 2025 (Fig. \ref{fig:stat}(b)). In terms of publication types, the corpus includes 19 full papers, 12 short papers, 12 posters, 6 abstracts, and 4 demos, reflecting a mix of mature contributions and emerging work in this space.

In the following sections, we organize our results around the four dimensions introduced in Section \ref{method:dimensions}. We first examine how accessibility is framed and who is positioned as the primary beneficiary of accessibility technologies. We then analyze how responsibility for accessibility is distributed across users, technologies, and other actors. Next, we investigate how DHH communities are conceptualized within the literature. Finally, we examine the spatial and perceptual assumptions that shape accessibility design in XR environments.

\subsection{Orientation toward Access}

We first report patterns in access framing and intended beneficiaries, and then turn to research practices that shape how access is studied and represented.

\subsubsection{Access Framing}
\label{result:dimension1:accessFraming}

Our initial coding scheme included three categories of accessibility approaches---\textit{translation}, \textit{sensory substitution}, and \textit{spatial reconfiguration}---which together captured the full corpus. Some studies did not fit neatly into a single category because they reflected multiple orientations. In these cases, we coded them based on the orientation most strongly emphasized in the paper's problem framing, design rationale, and reported contribution.

\begin{figure}[h]
  \centering
  \includegraphics[width=0.8\linewidth]{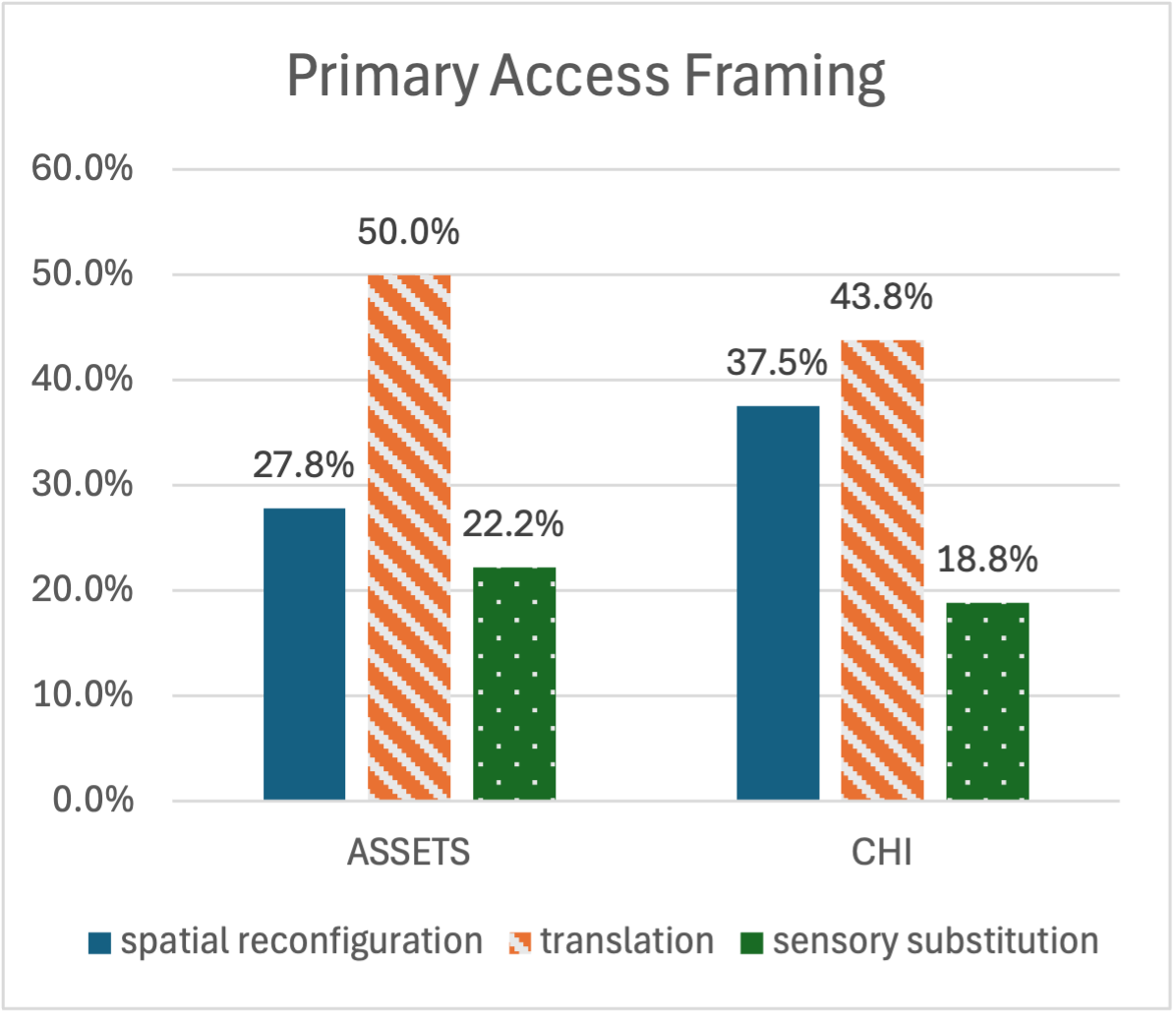}
  \caption{Comparison of primary access framing in XR accessibility research for DHH communities across ASSETS and CHI. Percentages are reported relative to the number of papers in each venue (ASSETS: $n=18$, CHI: $n=16$). Other venues are not included due to the smaller number of relevant papers in the corpus.}
  \Description{A grouped bar chart compares primary access framing across ASSETS and CHI. For ASSETS, spatial reconfiguration is 27.8\%, translation is 50.0\%, and sensory substitution is 22.2\%. For CHI, spatial reconfiguration is 37.5\%, translation is 43.8\%, and sensory substitution is 18.8\%.}
  \label{fig:stat1}
\end{figure}

Translation emerged as the dominant accessibility approach, accounting for $60.4\%$ ($32/53$) of the papers, with $17.0\%$ ($9/53$) focusing on sensory substitution and $22.6\%$ ($12/53$) on spatial reconfiguration. A similar overall pattern appears in both ASSETS and CHI papers (Fig. \ref{fig:stat1}). In most cases, accessibility is achieved by enabling communication between DHH users and hearing non-signing interlocutors through linguistic mediation. Many systems focus on visualizing and positioning sign language interpreters within immersive environments~\cite{paudyalDAVEEDeafAccessible2019, luoAvatarInterpreterImproving2022, andertonInvestigatingSignLanguage2022, yangHolographicSignLanguage2022}. Others implement captioning systems that overlay spoken language as text within the user's visual field~\cite{daviduburAIDrivenInterpretationNonverbal2025, xieVRCaptionsDesignCaptions2025, yamamotoSeeThroughCaptionsRealTime2021, jainAccessibleConversationsMobile2018, olwalWearableSubtitlesAugmenting2020}. Some studies explore automated translation approaches, such as converting speech into Signed Exact English~\cite{yangHolographicSignLanguage2022}, or developing two-way communication systems between DHH and hearing users~\cite{pengPosterSiHearSigntoHear2025}. Researchers have also experimented with alternative caption presentation formats, including speech balloons~\cite{zempoSpeechBalloonSystem2017} and speech bubbles~\cite{pengSpeechBubblesEnhancingCaptioning2018}. Recent papers also suggest that artificial intelligence (AI) is increasingly being incorporated into translation-oriented XR accessibility systems, particularly in captioning~\cite{daviduburAIDrivenInterpretationNonverbal2025}, sign language recognition~\cite{schioppoSignLanguageRecognition2019}, and automated translation~\cite{fanucchiFineTuningVideoMasked2024, caiSignGlassFirstPersonView2025, guoAssistiveARSystem2025}.

\subsubsection{Beneficiary Framing}

The intended beneficiaries of XR accessibility systems are usually specified across the ACM corpus. DHH users are a common target group, particularly in studies of sound awareness technologies that allow users to perceive auditory information in their surroundings~\cite{jainHeadMountedDisplayVisualizations2015, jainAccessibleConversationsMobile2018, SoundVizVRSoundIndicators, caoSoundModVRSoundModifications2024a}. Hearing novice signers are another recurring beneficiary group, especially in work that uses immersive environments to support ASL learning and practice~\cite{alamInsightsImmersiveLearning2024, quandtTeachingASLSigns2020, hossainSupportingASLCommunication2023, hossainContextresponsiveASLRecommendation2022}. XR developers also appear as intended beneficiaries in papers that provide development tools or taxonomies for integrating accessibility features into XR applications~\cite{jainTaxonomySoundsVirtual2021, segalSocialCueSwitchCustomizableAccessibility2024, fengToozKitSystemExperimenting2023, caoSoundModVRSoundModifications2024a}. Beyond these groups, Peng et al.~\cite{pengSpeechBubblesEnhancingCaptioning2018} observe that captioning interfaces designed to support DHH users may also assist hearing individuals in situations involving unfamiliar spoken languages, and suggest that accessibility technologies developed for DHH users can also benefit hearing users in some contexts.

Communicative benefits are not always evenly received. Systems described as being designed for DHH users may nevertheless primarily facilitate communication for hearing non-signing interlocutors rather than support reciprocal interaction (e.g.~\cite{fanucchiFineTuningVideoMasked2024, yamamotoSeeThroughCaptionsRealTime2021, tuConversationalGreetingDetection2020, yangHolographicSignLanguage2022}). For example, \textit{SignGlass} translates ASL into English without supporting communication in the reverse direction~\cite{caiSignGlassFirstPersonView2025}. In such cases, we interpret accessibility as remaining organized around hearing norms, with communicative benefits asymmetrically distributed.

\subsubsection{Research Practices}
\label{result:dimension1:methodologicalFraming}

\begin{figure}[h]
  \centering
  \includegraphics[width=0.8\linewidth]{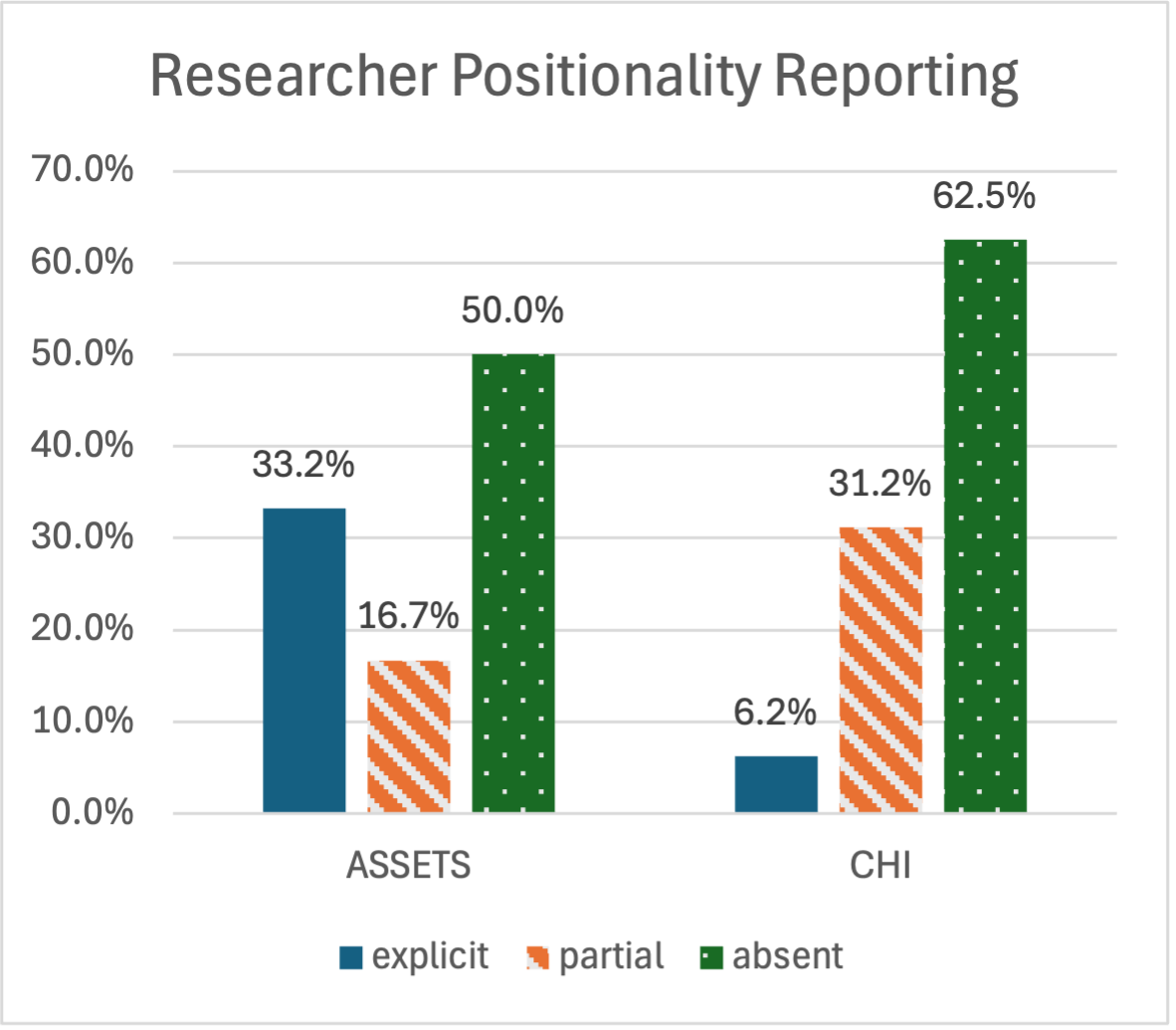}
  \caption{Comparison of researcher positionality reporting in XR accessibility research for DHH communities across ASSETS and CHI. Percentages are reported relative to the number of papers in each venue (ASSETS: $n=18$, CHI: $n=16$). Other venues are not included due to the smaller number of relevant papers in the corpus.}
  \Description{A grouped bar chart compares researcher positionality reporting across ASSETS and CHI. For ASSETS, explicit reporting is 33.3\%, partial reporting is 16.7\%, and absent reporting is 50.0\%. For CHI, explicit reporting is 6.2\%, partial reporting is 31.2\%, and absent reporting is 62.5\%.}
  \label{fig:stat2}
\end{figure}

Research practices varied in how studies situated researchers in relation to DHH communities and how they supported communication with DHH participants. Only $22.6\%$ ($12/53$) of papers explicitly reported the researchers' identity (e.g., hearing, Deaf, deaf, or hard of hearing), sign language fluency, or their relationship with DHH communities. An additional $17.0\%$ ($9/53$) briefly noted that their research teams included DHH members, but did not elaborate on team composition or how collaboration within the team was structured. Among the two major venues in our corpus, ASSETS demonstrated greater overall disclosure of researcher positionality than CHI (50.0\% vs. 37.5\%), particularly in the proportion of papers that included explicit positionality statements (33.3\% vs. 6.2\%; Fig. \ref{fig:stat2}).

Five papers in the corpus ($9.4\%$, $5/53$) explicitly described hearing researchers' positionality and connections with DHH communities~\cite{lunaCommunicationCollaborationCoordination2024, lunaExploringDeafHard2025, almutairiAugmentedRealityLiteracy2017, guoAssistiveARSystem2025, yamamotoSeeThroughCaptionsRealTime2021}. Luna et al. note that \textit{``As researchers, we acknowledge our position as outsiders to the DHH community... our interpretation of the results may be biased due to our position as hearing individuals''} and emphasize grounding their research in DHH participants' experiences~\cite{lunaCommunicationCollaborationCoordination2024, lunaExploringDeafHard2025}. 

Notably, a few papers led by DHH scholars explicitly reflected on how lived experience shaped the problem framing and research direction. These studies often situate the work within the authors' own experiences navigating accessibility barriers and DHH communication contexts (e.g.,~\cite{jainHeadMountedDisplayVisualizations2015, mathewAccessDemandRealtime2022, berkeChatHatPortable2020, jainAccessibleConversationsMobile2018}).

\subsection{Distribution of Responsibility for Access}

This dimension examines responsibility both in how XR systems are designed and used, and in how access is supported during research practices involving DHH participants.

\subsubsection{Responsibility Allocation}
\label{result:dimension2:responsibilityFraming}

In the design and use of XR systems, accessibility is commonly achieved by requiring DHH users to adopt additional tools, monitor translated information, or manage communication breakdowns within hearing-majority environments (e.g.~\cite{caiSignGlassFirstPersonView2025, millerUseSmartGlasses2017, yangHolographicSignLanguage2022, fanucchiFineTuningVideoMasked2024, luoAvatarInterpreterImproving2022}). Some papers explicitly acknowledge this tension, noting that accessibility often demands additional effort from DHH users, such as wearing heavy devices for extended periods of time (e.g. \cite{jainAccessibleConversationsMobile2018, jainExploringAugmentedReality2018}). Matthew et al.~\cite{mathewAccessDemandRealtime2022} and Berke et al.~\cite{berkeChatHatPortable2020} similarly describe situations in which DHH users must actively manage accessibility in everyday interactions.

At the same time, other studies treat accessibility as a form of shared labor by redistributing responsibility across sociotechnical systems. Some projects shift responsibility toward institutions or infrastructures, such as providing real-time sign language narration for live planetarium shows~\cite{jonesDeliveringSignLanguage2019} or designing mid-air haptic experiences in public exhibits for visitors with diverse accessibility needs~\cite{oconaillImprovingImmersiveExperiences2020}. Other studies shift responsibility toward system designers by offering taxonomies or toolkits that support the integration of accessibility features into XR applications~\cite{jainTaxonomySoundsVirtual2021, segalSocialCueSwitchCustomizableAccessibility2024, fengToozKitSystemExperimenting2023, caoSoundModVRSoundModifications2024a}. For example, Jain et al.~\cite{jainTaxonomySoundsVirtual2021} propose a taxonomy of sounds in virtual reality intended to help designers identify and incorporate sound accessibility considerations during system design. Accessibility labor is also redistributed across participants in the communication ecosystem, for example by involving parents, teachers, and interpreters in supporting Deaf children's language development~\cite{almutairiAugmentedRealityLiteracy2017, hossainContextresponsiveASLRecommendation2022, hossainSupportingASLCommunication2023}. Some studies also redistribute part of the accessibility labor to hearing interlocutors. For example, speakers may be prompted to adjust their speech behavior in response to feedback about caption quality~\cite{kangImprovingRealTimeHeadWorn2024}. Together, these approaches show that responsibility for access can be distributed across systems, institutions, and communication partners, rather than remaining individualized onto DHH users.

\subsubsection{Methodological Practices}
\label{result:dimension2:methodologicalPractices}

\begin{figure}[h]
  \centering
  \includegraphics[width=0.8\linewidth]{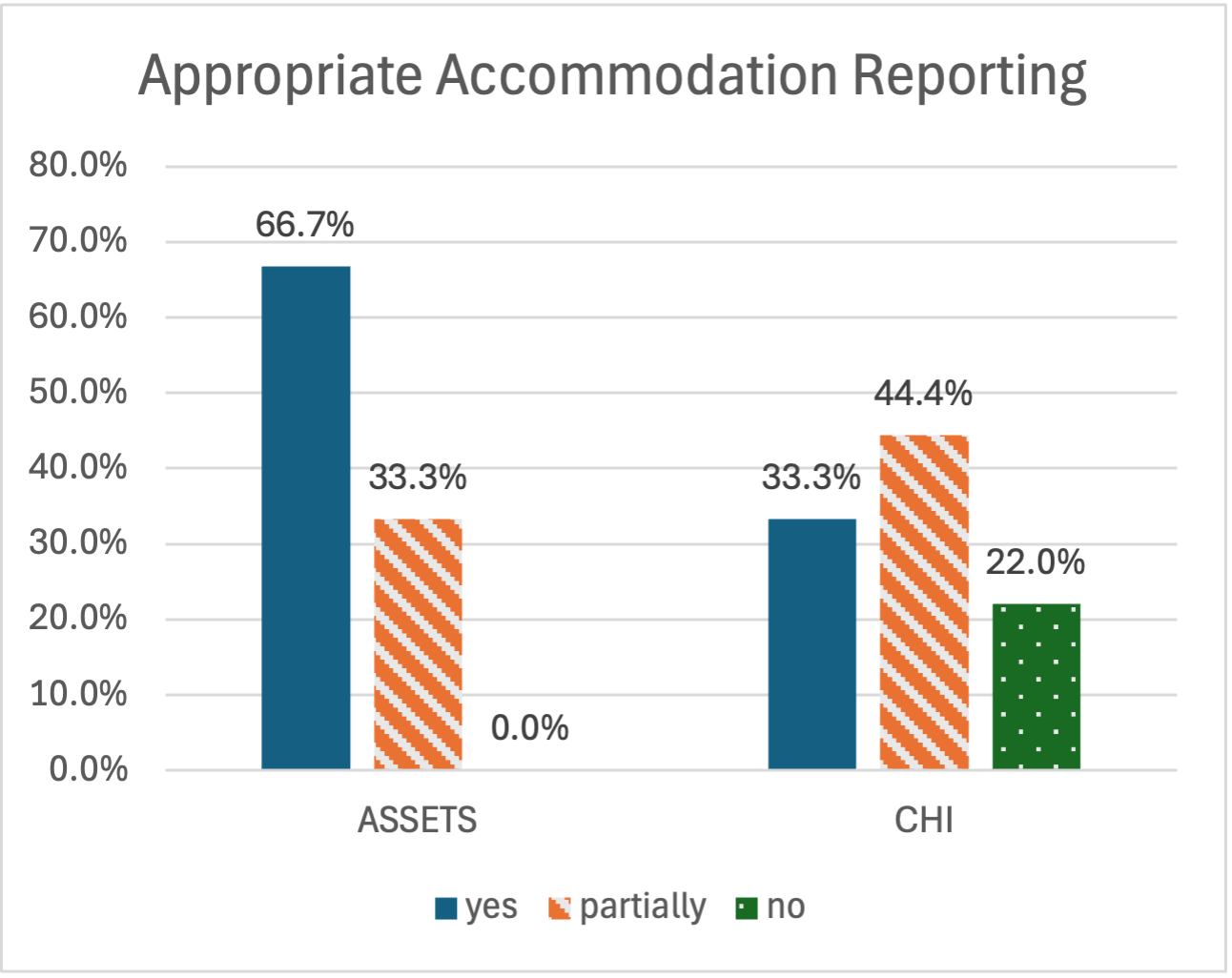}
  \caption{Comparison of appropriate accommodation reporting in XR accessibility research for DHH communities across ASSETS and CHI. Percentages are reported relative to the number of papers in each venue (ASSETS: $n=18$, CHI: $n=16$). Other venues are not included due to the smaller number of relevant papers in the corpus.}
  \Description{A grouped bar chart compares appropriate accommodation reporting across ASSETS and CHI. For ASSETS, yes is 66.7\%, partially is 33.3\%, and no is 0\%. For CHI, yes is 33.3\%, partially is 44.4\%, and no is 22.2\%.}
  \label{fig:stat3}
\end{figure}

Beyond researcher positionality discussed in Section \ref{result:dimension1:methodologicalFraming}, methodological practices also varied in how studies supported communication with DHH participants. Among the 39\footnote{Some papers did not involve user studies with DHH participants due to the nature of the research, such as work on sound taxonomies~\cite{jainTaxonomySoundsVirtual2021}. Others were published as posters or demos and did not yet include user studies, although later full papers reported such studies (e.g.,~\cite{caoSoundModVRSoundModifications2024, caoSupportingSoundAccessibility2024}). Therefore, in this section we only consider papers that explicitly reported conducting user studies with DHH participants.} papers that reported conducting user studies with DHH participants, 35.9\% (14/39) reported providing communication accommodations such as sign language interpreters or human-generated real-time captions, also known as Communication Access Real-time Translation (CART). No paper explicitly reported that researchers communicated directly with participants in sign language without such mediation. The remaining studies instead relied on alternative strategies that placed much of the responsibility for communication on DHH participants, such as lipreading, reading and monitoring speech-to-text output, written communication, or gestures, or did not report any accommodations or communication modalities. Among the two major venues, ASSETS demonstrated a higher rate of appropriate accommodation than CHI. All ASSETS papers reported accommodations in their user studies, either fully (66.7\%) or partially (33.3\%), whereas 22.2\% of CHI papers did not report any accommodation (Fig. \ref{fig:stat3}).

Vinayagamoorthy and Ziegler~\cite{vinayagamoorthyPersonalisingTVExperience2019} reflect on how communication arrangements during the study shaped the data collected. Describing differences across sites, they note: \textit{``there were at least two experimenters and a BSL\footnote{BSL: British Sign Language.} interpreter to help conduct studies in the UK while in Germany we had one experimenter who used the written form of German for communicating with participants. This might account for the more verbose responses from our UK participants.''} They further recommend \textit{``the use of a sign language interpreter if working with participants in the deaf community.''}

\subsection{Conceptualization of DHH Communities}
\label{result:dimension3}

\subsubsection{Dominant Conceptualizations}

Across the ACM corpus, deafness is understood in two primary ways: as a sensory impairment within a medical framing, and as a relatively homogeneous user group. First, in some studies, accessibility is approached through compensating for the absence of auditory perception. Serafin et al.~\cite{serafinListenAgainVirtual2023} present a VR-based training system for children with \textit{``hearing impairments,''} describing auditory challenges in deficit-oriented terms. The authors note that \textit{``factors such as distance to the sound source and noise from the surroundings are the children's biggest enemies''}. In other cases, communication differences are framed primarily in terms of individual capability. Yamamoto et al.~\cite{yamamotoSeeThroughCaptionsRealTime2021} state that \textit{``among DHH people, each person has his or her own preference for communication methods. Especially some people are \textbf{unable} to speak well and can \textbf{only} use the sign language.''} Some studies also situate DHH users within broader disability categories by designing systems intended to support multiple disabilities simultaneously, such as XR experiences for DHH, BLV~\cite{segalSocialCueSwitchCustomizableAccessibility2024}, and wheelchair users~\cite{oconaillImprovingImmersiveExperiences2020}. 

Second, DHH users are often treated as a relatively homogeneous group with similar needs and communication practices. Even when DHH participants are included in user studies, assumptions about uniform preferences appear. For example, some studies assume that captioning benefits all DHH users~\cite{segalSocialCueSwitchCustomizableAccessibility2024, tuConversationalGreetingDetection2020}, that all DHH people are unable to hear~\cite{tuConversationalGreetingDetection2020}, that all DHH users have limited reading or caption comprehension abilities~\cite{luoAvatarInterpreterImproving2022}, or that all DHH users use sign language~\cite{luoAvatarInterpreterImproving2022, paudyalDAVEEDeafAccessible2019, yangHolographicSignLanguage2022}.

Other research adopts a more nuanced view of DHH communities and their communication practices. Findlater et al. show that preferences for sound awareness technologies vary across DHH users and are shaped by multiple factors, such as life experience~\cite{findlaterDeafHardofhearingIndividuals2019}. Jain et al. note that sound awareness tools may only serve a subset of the DHH community, recognizing diversity within the population and discussing Deaf culture in relation to accessibility design~\cite{jainTaxonomySoundsVirtual2021}. Across a series of recent studies, Luna et al. examine communication practices in collaborative AR environments~\cite{lunaCommunicationCollaborationCoordination2024}, visual scanning behaviors in AR~\cite{lunaExploringVisualScanning2024}, and DHH perspectives on AR tasks~\cite{lunaExploringDeafHard2025}. Mathew et al. report that accessibility experiences with Apple Vision Pro are shaped by the spatial layout and interaction demands of the device~\cite{mathewIntheWildAccessibilityEvaluation2025}, while Britain et al. show that DHH users have distinct preferences for how captions are positioned and presented on head-worn displays in group conversations~\cite{britainPreferencesCaptioningEmulated2022}. Other studies consider the social dimensions of accessibility, such as concerns about stigmatizing or conspicuous assistive technologies~\cite{millerUseSmartGlasses2017}, or how disability disclosure and representation are negotiated in XR~\cite{zhangItsJustPart2022}. Taken together, these studies show that the corpus also includes work that treats DHH communities as diverse groups with different cultural backgrounds, communication practices, and accessibility preferences, rather than as a uniform accessibility category.

\subsubsection{Methodological Assumptions}

Methodological choices in the literature also reflect different conceptualizations of DHH users and their roles in the research process.

Some studies involve DHH participants primarily to evaluate system performance, with less attention to understanding how DHH users experience the system. For example, participants may be asked to complete predefined tasks and report performance measures such as accuracy, response time, or recognition rate, without examining how they interpret, adapt to, or make sense of the system in everyday use contexts~\cite{chelladuraiSoundHapticVRHeadBasedSpatial2024, dzardanovaSignLanguageImmersive2022, SoundVizVRSoundIndicators}. In other cases, systems designed for sign language interaction are evaluated by hearing people, such as sign language teachers, rather than by Deaf signers themselves~\cite{dzardanovaSignLanguageImmersive2022}. In some studies, DHH participants are involved only at later stages of the research process, such as providing feedback on already developed systems~\cite{yangHolographicSignLanguage2022, chelladuraiSoundHapticVRHeadBasedSpatial2024, SoundVizVRSoundIndicators, segalSocialCueSwitchCustomizableAccessibility2024, pengPosterSiHearSigntoHear2025, zempoSpeechBalloonSystem2017}. 

Oversimplified understandings can also shape study procedures. Participants are asked to remove hearing aids in order to \textit{``ensure the visualization was the focus in this study''}~\cite{SoundVizVRSoundIndicators} or to \textit{``eliminate auditory bias during the task''}~\cite{chelladuraiSoundHapticVRHeadBasedSpatial2024}, reflecting a limited understanding of DHH participants' diverse relationships to residual hearing and assistive devices. Such practices do not establish a neutral baseline, but instead introduce artificial conditions that disrupt participants' everyday access strategies. These examples suggest that assistive devices are treated as sources of bias to be controlled, rather than as integral components of participants' everyday life. Related assumptions also appear when hearing participants wearing noise-canceling headphones are used to emulate DHH users in experiments for systems intended \textit{``for people who are deaf and hard of hearing''}~\cite{tuConversationalGreetingDetection2020}, suggesting that deafness is understood solely as the inability to hear.

In contrast, some studies explicitly report recruiting DHH individuals with a range of communication methods and assistive technology experiences in order to capture diverse perspectives~\cite{findlaterDeafHardofhearingIndividuals2019, olwalWearableSubtitlesAugmenting2020, olwalWearableSubtitlesAugmenting2020}. Almutairi and Al-Megren also involve additional stakeholders such as parents, teachers, and interpreters when designing technologies for Deaf children~\cite{almutairiAugmentedRealityLiteracy2017}. Others engage extensively with DHH communities through user-centered, participatory, or co-design approaches (e.g.~\cite{samaradivakaraSeEarTailoringRealtime2024, caiSignGlassFirstPersonView2025, xieVRCaptionsDesignCaptions2025}).

\subsection{Spatial and Perceptual Assumptions}
\label{result:dimension4}

XR accessibility systems for DHH often consider the visual and spatial characteristics of signed communication, illustrating how immersive environments shape accessibility. As explained in Section \ref{rw:DeafSpace}, signed interaction relies on visual attention, clear sightlines, and coordinated spatial positioning among communicators. These considerations appear in studies examining interpreter placement~\cite{andertonInvestigatingSignLanguage2022, vinayagamoorthyPersonalisingTVExperience2019} and sightlines within immersive environments~\cite{luoAvatarInterpreterImproving2022, hossainSupportingASLCommunication2023}. Berke et al. address practical aspects of signed interaction by designing portable interpreter tools that allow users to keep both hands free during communication~\cite{berkeChatHatPortable2020}. Research on collaborative AR environments further explores how DHH users coordinate attention and communication in shared spaces~\cite{lunaCommunicationCollaborationCoordination2024, lunaExploringDeafHard2025}.

In many of these systems, the surrounding environment remains structured around audio-default practices, particularly in how attention, interaction, and information flow are organized. Accessibility features are often introduced to make auditory events visible or to support interaction with non-signing hearing users, for example through sound awareness~\cite{lunaExploringDeafHard2025, jainHeadMountedDisplayVisualizations2015}, sound localization~\cite{chelladuraiSoundHapticVRHeadBasedSpatial2024}, or sound recognition systems that inform users about auditory activity in their surroundings~\cite{guoHoloSoundCombiningSpeech2020, SoundVizVRSoundIndicators}. 


In these systems, spatial configurations of interaction typically remain unchanged, and accessibility is achieved by adding additional layers of information onto existing environments. We interpret this pattern as reflecting a hearing-centered organization of communication and attention, where accessibility is added to an audio-default environment rather than used to reconfigure the environment itself.

XR technologies are however fundamentally spatial media that allow environments to be dynamically configured. This spatial flexibility is increasingly recognized as a potential resource for accessibility design. Alam et al. propose immersive environments as a way to simulate real-life interactions for sign language learning~\cite{alamInsightsImmersiveLearning2024}. Mathew et al. examine accessibility experiences with emerging XR devices and highlight how spatial layout influences usability in immersive contexts~\cite{mathewIntheWildAccessibilityEvaluation2025}. Work on XR captioning similarly demonstrates how spatial positioning and layout affect the effectiveness of accessibility features in immersive environments~\cite{xieVRCaptionsDesignCaptions2025, glasserMixedRealitySpeaker2019}. 

Rather than treating XR as an augmented hearing-default environment, a smaller body of work explores how immersive environments can be organized around Deaf ways of being. Chateauvert et al.~\cite{chateauvertRemixingFlyingWords2025} experiment with creative expressions of ASL poetry, treating sign language as a spatial and expressive medium. Similarly, Jones and Lawler~\cite{jonesDeliveringSignLanguage2019} design a system that uses HMDs and infrared tracking to provide real-time sign language narration for live planetarium shows, integrating signed communication into the viewing experience. Together, these studies illustrate how XR can be used not only to augment hearing-default environments, but also to organize immersive spaces around DHH communication practices.

\section{Discussion}

Our analysis of this ACM corpus suggests that when accessibility research frames deafness as a communication deficit, it shapes how responsibility for access is distributed and constrains how XR accessibility is imagined, often within hearing-default terms. In this discussion, we use our theoretical framework to interpret these relationships and to consider how it might orient more generative approaches to XR accessibility research and design. We first reconsider how DHH communities are conceptualized beyond deficit-oriented framings, then examine how accessibility responsibilities become individualized onto DHH users and how they might instead be negotiated and distributed across sociotechnical systems. Finally, we turn to XR as a spatial medium, exploring how its affordances open possibilities for reimagining accessibility toward alternative imaginaries grounded in Deaf epistemologies.

\subsection{Communication Equity: Beyond Communication Access}

Communication barriers remain a central access issue for many DHH people, especially in mixed-hearing settings where participants may not share the same language, modality, or communication norms. Research on captions, interpreting, translation, and other forms of communication support therefore remains important. Consistent with prior reviews~\cite{mack2021we}, our analysis of this selected ACM corpus shows a recurring tendency for DHH accessibility in XR to be framed primarily through communication mediation. Within this framing, access can appear to be addressed once speech, environmental sound, or sign language has been translated into another modality. This narrows the design space for XR accessibility. DHH users may be positioned primarily as individuals who require technological mediation to communicate with hearing people, while other dimensions of equitable access---such as spatial organization, cultural and linguistic specificity, and the distribution of accessibility labor---receive less attention.

\begin{figure}[h]
  \centering
  \includegraphics[width=0.8\linewidth]{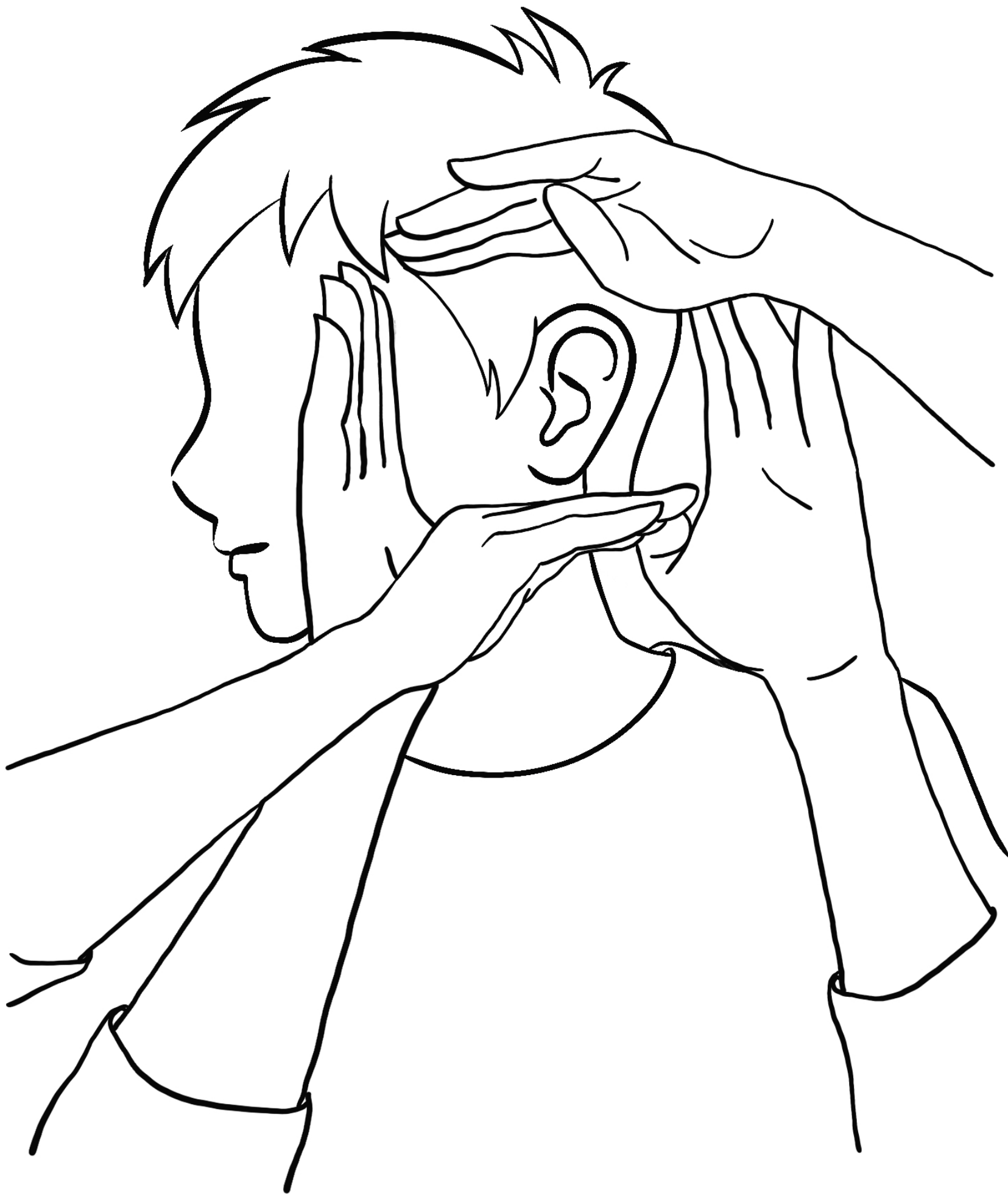}
  \caption{The ASL sign for AUDISM, glossed as EAR BOX. The sign conveys a Deaf understanding of audism as a hearing-centered frame that boxes communication, access, and human value into the ear through auditory norms~\cite{asl_university_audism_ear_box}.}
  \Description{An illustrated person demonstrates the ASL sign for AUDISM (glossed as EAR BOX) by forming a box shape around the ear with both hands, representing a hearing-centered, auditory-bounded frame.}
  \label{fig:audism}
\end{figure}

Here, we foreground a Deaf perspective through Fig.~\ref{fig:audism}. Audism, a foundational concept in \textit{Deaf Studies}, refers to a system of beliefs and practices that center hearing as the normative basis for communication and social value~\cite{bauman2004audism, humphries1975making}. The ASL sign for \textit{AUDISM}, glossed as \textit{EAR BOX}, visually represents this concept by enclosing the ear within a bounded space, conveying a Deaf understanding of audism as a hearing-centered, auditory-bounded frame.

Deafness is not inherently a communication deficit. \textit{Deaf Studies} scholars have argued that communication barriers are shaped by social organization rather than hearing status alone. The history of \textit{Martha's Vineyard} illustrates this point: in a Massachusetts community with a high prevalence of hereditary deafness, both hearing and DHH residents used sign language in everyday life. In that context, deafness did not carry the same communication barriers that arise in communities organized primarily around spoken language~\cite{groce1985everyone, power2024historical}. This example shows that spoken and signed languages are different modalities rather than hierarchically ordered forms of communication.

HCI scholars have similarly emphasized that communication norms are socially constructed and, therefore, designable~\cite{mcdonnell2024envisioning, mcdonnell2023easier}. Moreover, Deaf communities have long adapted and repurposed technologies for their own communication practices. Technologies such as fax machines~\cite{communications_access_deaf_eyes}, text messaging~\cite{notepage_text_messaging_deaf}, the Sidekick phone~\cite{retromobe_sidekick_lx_2009}, and more recently LLM tools~\cite{huffman2025we} have been widely adopted and creatively used within DHH communities, sometimes influencing communication practices beyond those communities~\cite{bitman2019deaf}. Through these histories, Deaf communities challenge the framing of DHH users as passive recipients of assistive technologies and instead highlight their role as active contributors in shaping technological practices.

\paragraph{Design Implications: From Impairment to Community} 

Connecting to our theoretical framework, the dimension of \textit{conceptualization of DHH communities} highlights that how DHH communities are understood is foundational to XR accessibility research. In particular, it shapes \textit{the orientation toward access}: how access is defined, which problems are prioritized, and what kinds of solutions are imagined. When DHH users are framed primarily through impairment, accessibility is often approached as a matter of compensating for hearing loss. In XR research, this framing can narrow design attention toward communication aids such as captioning and interpretation, rather than toward richer engagement with virtual environments themselves. As a result, the field risks treating XR primarily as a tool for ``repairing'' communication with hearing systems, rather than as a medium for learning, workplace training, gaming, collaboration, and other forms of interaction on their own terms.

In contrast, understanding DHH communities as linguistic and cultural groups shifts the focus toward supporting diverse communicative practices. Designing from this perspective also requires avoiding a single default profile of the DHH user. XR access needs may vary across language use, hearing status, assistive technology use, communication partners, and task context. For some users and situations, captions may be central; for others, access may depend more on signer visibility and ASL-first spatial organization. Users with residual hearing and hearing-assistive devices such as hearing aids and cochlear implants may also need headset designs that account for device fit and comfort. In mixed-hearing collaboration, access may depend on whether communication partners sign, speak, rely on captions, or move across multiple modalities. These differences are not mutually exclusive categories, since DHH users often combine multiple strategies including signing, captions, residual hearing, hearing technologies, written communication, and environmental cues across contexts. XR systems should therefore support configurable access ecologies, such as user control over caption placement and speaker attribution; flexible signer-content layouts that preserve sightlines; visual or haptic cues for attention repair; and modality choices that can be adjusted across tasks and communication partners.

These concerns also apply to study design. As discussed in Section~\ref{result:dimension3}, studies involving deeper participation or leadership from DHH participants and researchers tend to reflect more nuanced understandings of both DHH communities and accessibility design. Prior reviews of HCI accessibility research similarly emphasize participatory approaches that center disabled users in the design process~\cite{alonzo2025review}. Assistive technologies should be treated as part of participants' everyday access configurations. Hearing aids, cochlear implants, and preferred communication arrangements should generally remain available during XR studies. If a study restricts any of these configurations, the restriction should be tied to a specific research question and discussed with participants. When access configurations may shape study outcomes, researchers can address this through documentation and analysis instead of removing them by default. For example, researchers can report participants' hearing technology use, captioning or signing preferences, headset fit, and any changes made during the study. These factors can then be considered when interpreting results, such as by reporting subgroup patterns, qualitative differences, or cases where access configurations shaped performance, comfort, or interaction. This approach treats DHH diversity as methodologically meaningful rather than as noise to be controlled away.

We therefore encourage greater transparency regarding researchers' positionalities, language abilities, and relationships with DHH communities\footnote{We recognize that positionality disclosure can be sensitive and may not be equally safe, comfortable, or appropriate for all researchers. Rather than prescribing a single form of disclosure, we emphasize the value of providing contextual reflection to help readers understand how researchers' relationships to DHH communities may influence the work.}, as well as methodological approaches that enable deeper engagement with DHH perspectives, including qualitative methods and participatory design conducted with appropriate accommodations, such as sign language interpreters and CART~\cite{harris2009research, ada_effective_communication}. We also note the absence of explicit reporting on direct sign language communication between researchers and DHH participants in the studies we reviewed. Even when interpreters, CART, or other communication arrangements are used, we encourage researchers to report these choices explicitly, as they shape both participation and the kinds of knowledge produced through the research process~\cite{singleton2014toward}.

\subsection{Access as Shared Responsibility}

Although interaction with computing systems necessarily involves all users, responsibility for accessibility is not evenly distributed~\cite{power2023mapping}. Rather, accessibility work is often implicitly allocated in ways that require DHH users to carry disproportionate communicative, cognitive, and technical labor~\cite{mcdonnell2022understanding, mcdonnell2023easier}. This includes both the use of accessibility features and the ongoing work of configuring systems, monitoring multiple information streams, and managing communication breakdowns~\cite{mack2025modeling}. Such burdens may be further compounded by embodied and device-level challenges. For instance, we found no papers in our corpus that explicitly addressed how hearing devices may interfere with XR headsets, despite the fact that this can substantially shape access in practice. We note this absence with particular concern given our own experiences as Deaf researchers using XR devices, as well as prior reports documenting similar challenges~\cite{huffman2026reclaiming, voicesofvr_xr_accessibility_dhh}. 

HCI scholars have argued that technologies do not merely respond to disability, but actively participate in shaping how disability is understood and addressed~\cite{spiel2022adhd, spiel2019agency}. This perspective suggests that accessibility is not a neutral requirement of interaction, but reflects broader assumptions about who is expected to adapt. While all users engage with tools, accessibility work is often naturalized as the responsibility of DHH users, positioning them as the primary managers of access. This asymmetry is also evident in research practices: the absence of appropriate accommodations can increase the cognitive effort required for DHH participants, constrain how they express themselves, and shape the kinds of data that are collected. Over time, such practices risk producing distorted knowledge, reinforcing unequal power relations, and eroding trust between DHH participants and hearing researchers~\cite{harris2009research}.

\paragraph{Design Implications: Redistributing Responsibility through Technology} 

\textit{Deaf Studies} and \textit{Disability Studies} scholarship suggest that access can be structurally distributed across users, institutions, and infrastructures~\cite{goering2015rethinking, ladd2005deafhood, groce1985everyone, lane2002deaf}. HCI researchers have similarly emphasized that communication responsibilities can be shared across all parties rather than individualized onto DHH users~\cite{mcdonnell2023easier, mcdonnell2022understanding}. As discussed in Section \ref{result:dimension2:responsibilityFraming}, this possibility is already visible in our corpus: responsibility for access is redistributed through institutions and infrastructures, such as live sign language narration in planetarium shows and accessibility design in public exhibits, as well as system designers and broader communication ecosystems.

However, redistributing accessibility labor is not without tension. Kang et al.~\cite{kangImprovingRealTimeHeadWorn2024} found that several hearing participants preferred not to use a system that provided visual feedback on caption quality because it required additional effort on their part, despite recognizing its overall benefits for their DHH interlocutors. This suggests that redistributing communicative responsibility may collide with established \textit{cultural expectations} about who is expected to adapt in mixed-hearing interactions. At the same time, such friction is not necessarily a sign of failure: technologies do not simply accommodate existing norms; they can also unsettle and reshape them by making alternative distributions of access possible~\cite{hamraie2017building, winner2017artifacts}.

Connecting to our theoretical framework, the four dimensions can be translated into questions for future XR research. \textit{Orientation toward access} asks how access is being conceptualized---as translation, substitution, or spatial reconfiguration. \textit{Distribution of responsibility} asks who is expected to carry the work of maintaining access. \textit{Conceptualization of DHH communities} asks which communities are being addressed, and whether they are treated as distinct or collapsed into a single accessibility category. Finally, \textit{spatial and perceptual assumptions of XR systems} asks whether accessibility solutions are simply added as overlays, or whether space, attention, gaze, and interaction flow are being fundamentally reorganized.

We therefore call for reframing accessibility: not as bringing DHH users into hearing systems, but as designing systems and interactional arrangements that can support different ways of being. XR technologies, as space-building systems, have the potential to negotiate and redistribute accessibility labor at scale. By recognizing accessibility as shared labor and designing for its redistribution, the field can move beyond hearing-centered assumptions and toward more inclusive and generative forms of access.

\subsection{(Re)Imagining Access through Deaf Possibilities}

\begin{figure*}[h]
  \centering
  \includegraphics[width=\linewidth]{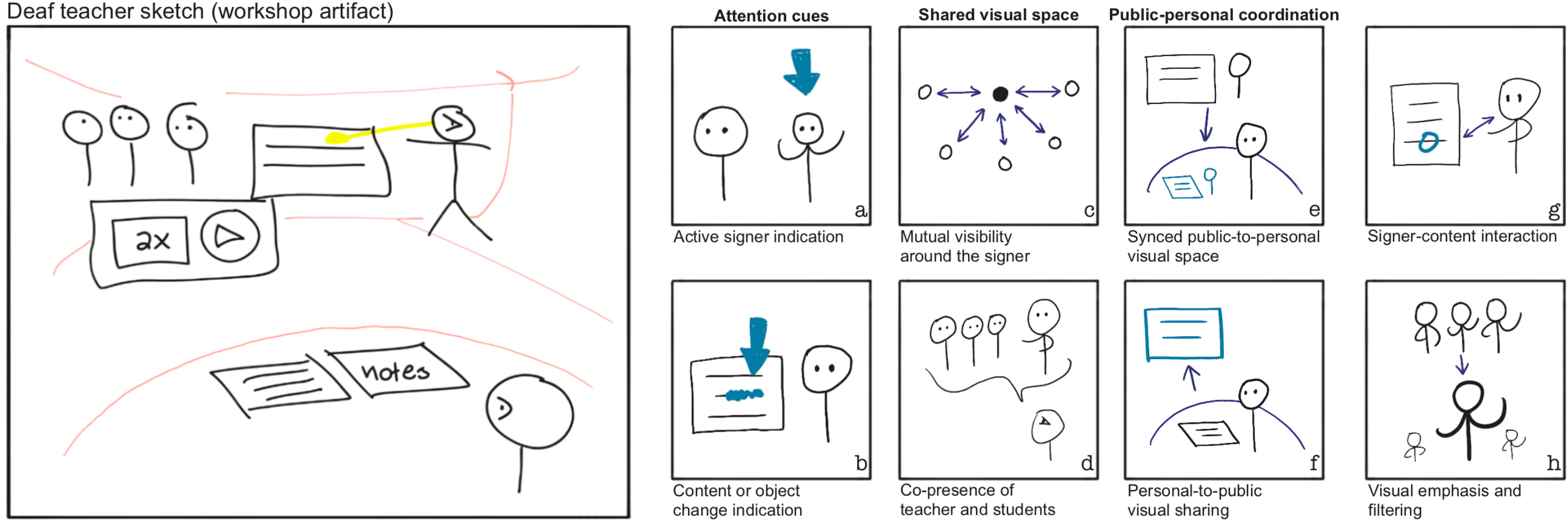}
  \caption{Illustrations of Deaf teachers' visions of Deaf-centered immersive classrooms. The left shows an original sketch created by a Deaf teacher. The right presents eight key design aspects presented in the paper: (a) active signer indication, (b) content or object change indication, (c) mutual visibility around the signer, (d) co-presence of teacher and students, (e) synced public-to-personal visual space, (f) personal-to-public visual sharing, (g) signer-content interaction, and (h) visual emphasis and filtering. Together, these elements foreground spatial coordination, shared visual access, and dynamic attention cues, illustrating how XR environments can be reconfigured to support Deaf ways of seeing and interacting. Adapted from Huffman et al.~\cite{huffman2026vision}, licensed under \href{https://creativecommons.org/licenses/by/4.0/}{CC BY 4.0}.}
  \Description{Composite figure with one large panel on the left and eight small panels on the right. The left panel is a hand drawn classroom sketch inside a rectangle: three student figures on the left face toward the center; a large rectangular board or screen with horizontal lines is in the middle; a teacher figure stands on the right pointing toward the screen, with a small yellow highlight on the screen near the pointing hand; in the lower left foreground is a large rectangular device showing ``2x'' and a play icon; in the lower middle are two sheets on a surface, one labeled “notes”; a head profile at the lower right faces toward the notes and the front of the room; curved red lines suggest sight lines or boundaries. The right side contains eight labeled mini diagrams (a–h) arranged in eight columns: (a) two figures with a large blue downward arrow pointing to the right figure (active signer indication); (b) a screen with a large blue downward arrow pointing to a changed or highlighted area (content or object change indication); (c) a central dot with surrounding dots and arrows indicating many-to-one visibility around a focal point (mutual visibility); (d) a teacher figure facing a group of student figures within a shared boundary line (co-presence); (e) a public board icon above a student figure with a downward arrow into a curved ``personal'' space (synced public-to-personal visual space); (f) a personal note icon near the student with an upward arrow to a public board icon (personal-to-public sharing); (g) a teacher figure pointing to a board with a highlighted mark and an arrow linking the teacher to the board (signer–content interaction); (h) a group of figures above with a downward arrow to a single larger central figure, with smaller figures around the central figure (visual emphasis and filtering).}
  \label{fig:demo}
\end{figure*}

\textit{Disability Studies} and HCI scholars have emphasized that accessibility research is not neutral, but embedded within broader political histories of disability that shape technological design~\cite{kafer2013feminist, mankoff_disability_2010}. The direction of the field reflects ongoing tensions between accommodation within existing norms and structural reconfiguration of those norms. In our corpus, we observe emerging work that begins to reconfigure space around Deaf epistemologies (Section~\ref{result:dimension4}). Related work beyond our corpus has also explored Deaf-centered XR futures through speculative workshops with Deaf teachers~\cite{huffman2026vision}, as illustrated in Fig. \ref{fig:demo}.

Concepts such as hearing \underline{loss}\footnote{While the notion of ``hearing loss'' is commonly used in disability, medical, and aging contexts, it does not uniformly apply to all DHH individuals. In particular, it may conflate distinct experiences such as age-related hearing decline and Deaf identity grounded in sign language and culture. We use the term here to reflect dominant deficit-oriented framings.} and Deaf \underline{Gain} illustrate how deafness can be understood either through deficit or through difference and possibility. Recognizing linguistic and cultural plurality moves immersive space design beyond correction and toward \textit{Deaf possibilities}.

\paragraph{Design Implications: Creating DeafSpace in XR} Building on the final dimension of our theoretical framework, \textit{spatial and perceptual assumptions}, we consider how access might be reimagined through Deaf epistemologies. XR technologies offer unique possibilities for designing interaction spaces. DHH communities have long developed ways of organizing visual communication through lived experience, which have informed \textit{DeafSpace}, an architectural approach that demonstrates how environments can support Deaf ways of seeing and interacting~\cite{edwards2014deafspace}. However, architectural environments remain constrained by physical structures. XR, as a space-building technology, is \textit{not} limited in the same way. Immersive environments can be dynamically configured, offering new opportunities to reconstruct interaction spaces. Designing for accessibility in XR therefore involves not only mediating communication between interlocutors, but also rethinking how space, attention, gaze, and interaction flow are structured.

We hope our theoretical framework offers a tool for transformation, one that moves XR research beyond reproducing hearing-default and deficit-oriented interaction patterns and toward creating spaces that support more diverse ways of connecting with one another and with the world. We imagine Deaf futures \textbf{bigger than the \textit{EAR BOX}}.

\subsection{Limitations}
\label{discussion:limitations}
Although the authors of this paper come from diverse geographic regions, including Asia, Europe, Africa, North America, and the Caribbean, the majority of the research team has been educated and/or employed in the United States. Moreover, our collaboration primarily took place through ASL and written English. As a result, our interpretations are inevitably shaped by our experiences with American Deaf culture. While we have attempted to engage with literature representing diverse contexts, our perspectives may still reflect assumptions and priorities common within U.S.-based Deaf communities. 

Additionally, our theoretical framework places particular emphasis on Deaf cultural, linguistic, and spatial perspectives, especially those related to signed and visually oriented communication. As such, the review is better suited to analyzing XR research that engages Deaf-centered communication practices than research involving DHH users who do not primarily associate with sign language, visual communication, or Deaf culture. Related issues, such as audio passthrough and hardware form factors, may also shape XR accessibility for these groups, but are not examined in depth in this paper. 

Finally, our corpus is bounded to selected ACM venues indexed in the ACM Digital Library. We chose this scope to examine a coherent ACM-published corpus within HCI, accessibility, and interactive computing, rather than to provide an exhaustive review of all XR scholarship. Similar ACM-bounded review approaches have been used in prior ASSETS work (e.g.,~\cite{mcdonnell2024envisioning, alonzo2025review}). Important work on XR accessibility is also published in adjacent XR venues and journals, including IEEE VR, ISMAR, TVCG, and other non-ACM research communities. As a result, our findings should be read as patterns within this ACM corpus, not as claims about the entire XR field. Applying our framework to broader XR literatures is an important direction for future work and may reveal additional framings, methods, and design priorities.

\section{Conclusion}

Drawing on \textit{Disability Studies}, \textit{Deaf Studies}, and \textit{DeafSpace}, we developed a theoretical framework and derived four analytical dimensions: orientation toward access, distribution of responsibility, conceptualization of DHH communities, and spatial and perceptual assumptions. We used this framework as a lens to analyze $53$ ACM-published XR studies from the past decade. Through this analysis, we identified recurring patterns in how access, users, and interaction are framed and practiced, offering a perspective on how XR accessibility has been collectively approached in the literature. By situating these patterns within our theoretical framework, we highlight opportunities to further explore how accessibility in XR might be conceptualized and designed in ways that better engage with DHH experiences and spatial ways of interaction. We hope this work contributes to ongoing discussions in the ASSETS community and supports future research in this space, moving toward the creation of \textit{DeafSpace in XR}.

\section*{AI Use Statement}
Generative AI was used as a reading aid by one coder to support keyword extraction from a portion of the dataset. It was not used to generate codes, develop themes, make analytical decisions, or write the manuscript. All coding decisions were made by the research team and validated through calibration, cross-review, and consensus processes. The authors take full responsibility for the analysis, interpretation, and writing.

\begin{acks}
We thank Paige DeVries for helping with data analysis. We thank Richard Ladner, Jennifer Mankoff, Aashaka Desai, Emma McDonnell, and Paige Foreman for their valuable feedback and support throughout this work. We also acknowledge the Deaf communities whose knowledge, practices, and critical perspectives have shaped the foundations of this research.

The contents of this paper were developed under a grant from the National Institute on Disability, Independent Living, and Rehabilitation Research (NIDILRR grant number 90REGE0027). NIDILRR is a Center within the Administration for Community Living (ACL), Department of Health and Human Services (HHS). The contents of this paper do not necessarily represent the policy of NIDILRR, ACL, or HHS, and endorsement by the Federal Government should not be assumed. This material is also based upon work supported by the National Science Foundation under Award Nos. 2425713 and 2348221. This work was partially funded by Equipex+ Continuum ANR-21-ESRE-0030 and supported by a French government grant managed by the Agence Nationale de la Recherche as part of the France 2030 program, reference ANR22-EXEN-0002 (PEPR eNSEMBLE / CATS). Shuxu Huffman was supported by an NSF CSGrad4US Graduate Fellowship.
\end{acks}

\bibliographystyle{ACM-Reference-Format}
\bibliography{refs}

\appendix

\end{document}